\documentclass[a4paper,12pt]{article}
\usepackage{amsmath,amssymb}
\usepackage[dvips]{graphics}
\usepackage[dviwin]{graphicx}

\begin{document}

\begin{center}
\section*{Infinite planar string:\\
cusps, braids and soliton exitations}

{S.V. Talalov}

\vspace{5 mm}

{\small Dept. of Theoretical Physics, Tolyatti State University, \\14 Belorusskaya str.,
Tolyatti Samara region 445667 RUSSIA \\ e-mail {svtalalov@tltsu.ru}}

\end{center}

{\bf keywords:} strings, braids, anyon models.

{\bf MSC codes:}  74K05,  81T30

\begin{abstract}
{~~~ We investigate   infinite strings in $(2+1)D$ space-time, which
may be considered  as   excitations of straight lines on the spatial
plane. We also propose the hamiltonian description of such objects
that differs from the standard hamiltonian description of the
string. The hamiltonian variables are separated into two independent
groups: the ''internal'' and ''external''
 variables.  The first ones are invariant under space-time transformations and are connected with the second form of the world-sheet.
  The ''external'' variables define the embedding of the world-sheet into space-time.
 The constructed phase space is nontrivial because the finite number
of constraints entangles the variables from these groups. First
group of the variables constitute the coefficients for the pair of
first-order spectral problems; the solution of these problems  is
necessary for the reconstruction of the string world-sheet.
 We consider the excitations,  which correspond to ''N- soliton'' solution of the spectral problem, and demonstrate that the
reconstructed string has cuspidal points. World lines of such points
form  braids of various topologies.}
\end{abstract}

\section{Introducton}
 \indent

   The interest for the dynamical systems on a plane
 was aroused in connection with the idea  of topological quantum calculations \cite{Kitaev}.
The planar string is the example of such system; we suppose that the
excitations of the string may be considered as some quasiparticles
with anyon statistics. Recently, the non-standard point of view on
elementary particles as the defects of the string condensed matter
(''string-net condensation'') was suggested \cite{LevWen}.
Indirectly, our approach corresponds to these ideas.

The cases of both closed and open string in arbitrary space-time
dimensions  were well-investigated in the literature (see, for
example, \cite{Zwi,GShW}). One of the frequently discussed objects
here is the first form {\bf I} of the world-sheet -- as opposed to
the second form {\bf II}. In this work we investigate  the
excitations of an infinite string on a spatial plane in terms of
second form {\bf II} and discuss a possible interpretation of such
excitations as some anyon-type quasiparticles. The complete
classical theory in terms of non-standard hamiltonian variables will
be constructed here.
 The suggested scheme is based on the geometrical approach \cite{TMF00} that is a generalization of a standard
 geometrical approach \cite{BarNes} in a string theory.
 Certain   ideas of this article can be founded in the work \cite{TMF07} and \cite{TMF09}.
 
 The considered strings can be viewed as embedded in four-dimensional space-time; that is why the main points
 of the  approach \cite{TMF07} developed for $4D$ relativistic string should be outlined.
 We start with the Polyakov action
 \begin{equation}
 S = -\frac{\gamma}{2} \int\sqrt{h}\,h^{ij}\partial_i X_\mu \partial_j X^\nu  d\xi^0d\xi^1,,
 \label{action}
 \end{equation}
where the following notations are introduced: $h^{ij}$ -- metric
tensor field in two-dimensional space-time $(\xi^0,\xi^1)$, $h =
\det(h^{ij})$ and
  $X_\mu(\xi^0,\xi^1)$ --
scalar fields in two-dimensional space-time with isotopic index
$\mu$. The isotopic space  is Minkowski space-time $E_{1,3}$ with
the metric tensor $g_{\mu\nu} = diag(1,-1,-1,-1)$. The standard
procedure \cite{GShW} leads to the equalities
$$\partial_{+}\partial_{-}X_\mu =0\,,\qquad  (\partial_{\pm}X_\mu)^2 = 0 \qquad\qquad
(\partial_\pm = \partial/\partial \xi_\pm,
\,\xi_\pm =\xi^1\pm\xi^0)\,, $$
 and so  the objects of our consideration will be time-like world-sheets with orthonormal parametrization.
 We consider the infinite strings, that's why we must impose
 asymptotic conditions on a curve $X_\mu = X_\mu(\xi^1)$ for any value of evolution parameter $\xi^0$.
 The demanded conditions  will be formulated  later; in the first place  we'll  investigate the local structure of the world-sheet.

 Let us define  the pair of light-like vectors in space $E_{1,3}$:
 \begin{equation}
 \label{e_pm}
 {\bf e}_{\pm}(\xi_\pm) =  \pm{1\over\varkappa} \,\partial_{\pm}{\bf X}(\xi_\pm)\,,
  \end{equation}
 where $\varkappa$ is an arbitrary positive constant.
 It is clear that we can  construct  the pair of orthonormal bases
 ${\bf e}_{\nu\pm}(\xi_\pm)$ that are connected with the introduced vectors  ${\bf e}_{\pm}$
 by the equalities
 ${\bf  e}_{\pm}=\left({\bf e_{0\pm}} \mp {\bf e_{3\pm}} \right)/2$.
 These bases allow us to define the vector-matrices  ${\bf\hat E_{\pm}}$:
 \begin{equation}
{\bf\hat E_{\pm}}= {\bf e_0}_\pm {\boldsymbol{1_2}} - \sum_{i=1}^3{\bf e_i}_{\pm}{\boldsymbol{\sigma_i}}\,.
     \label{matr_E}
\end{equation}
The stationary groups for the vectors ${\bf  e}_{\pm}$ will be three
- parameter groups. Consequently, the definition of the bases ${\bf
e}_{\nu\pm}(\xi_\pm)$ has three - parameter arbitrariness in each
point $(\xi^0,\xi^1)$. We will keep  this fact in mind and eliminate
this ambiguity in the next section. The important object for our
subsequent considerations  is the $SL(2,C)$ - valued field
$K(\xi^0,\xi^1)$.   This field  is defined as follows:

\begin{equation}
{\bf\hat E_+} = K{\bf\hat E_-}K^{+}\,.
     \label{field_K}
\end{equation}

To fulfill the reduction into D = 1+2 space-time  we must require
the matrix $K$ to be real, so that $K \in SL(2,R)$. This requirement
means that
$${\bf e_{2+}}(\xi_+) = {\bf e_{2-}}(\xi_-) = {\displaystyle\sf\bf b}_2\,,$$
where ${\displaystyle\sf\bf b}_2$ is a constant spatial vector
so that the reduced space-time
$E_{1,2} \perp {\bf b}_2$.

In accordance with the definition of the vector-matrices ${\bf\hat
E_\pm}$, these matrices satisfy the equalities $\partial_{\pm}
{\bf\hat E_\mp}=0$.
 As the consequence, the matrix field $K(\xi^0,\xi^1)$  satisfies  to (special) WZWN - equation
\begin{equation}
\partial_+\left(K^{-1}\partial_-K\right)=0\,.
     \label{eq_K}
\end{equation}
Let us define the real functions $\varphi(\xi^0,\xi^1)$ and
$\alpha_\pm(\xi^0,\xi^1)$ by means of Gauss decomposition for the
matrix $K(\xi^0,\xi^1)$:
\begin{equation}
K = \left(\begin{matrix}
1&0\\
{-\alpha_+}&1
\end{matrix}\right)
\left(\begin{matrix}\exp(-{\varphi}/{2})&0\\
0&\exp({\varphi /2})\end{matrix}\right)
\left(\begin{matrix}1&\alpha_-\\
 0&1\end{matrix}\right)\,.
     \label{gauss}
\end{equation}
In general, these functions   are singular because the decomposition
(\ref{gauss}) is not defined for the points where the principal
minor $K_{11}=0$. Let us introduce the regular functions
$\rho_{\pm}=(\partial_{\pm}\alpha_{\mp})\,{\exp}(-\varphi)$. The
consequence of the equality (\ref{eq_K}) will be  the following PDE
- system:
\begin{subequations}
\label{TL_rho}
     \begin{eqnarray}
     \label{1_sys}
     \partial_+\partial_-\varphi &=& 2\rho_+\rho_-\exp\varphi,\\
     \label{2_sys}
     \partial_\pm\rho_\mp &=& 0,\\
     \label{3_sys}
     \partial_\pm\alpha_\mp &=& \rho_\pm\exp\varphi.
     \end{eqnarray}
\end{subequations}
 The introduction of the function $\varphi$ and the functions $\rho_{\pm}$ is justified by the following formulae for the first  $({\bf I})$
and the second $({\bf II})$ forms of the world-sheet:
$$ {\bf I} =  -\frac{\varkappa^2}{2}\,{\rm e}^{-\varphi}d\xi_+d\xi_-\,,\qquad
{\bf II} = \varkappa [\rho_+ d\xi^{2}_+   -  \rho_- d\xi^{2}_- ]  \,. $$

The standard method of geometrical description of a string
\cite{BarNes} uses the equations (\ref{1_sys}) and (\ref{2_sys})
which are deduced from the Gauss and Peterson-Kodazzi equations. In
a standard approach the inequalities $\rho_\mp > 0$ are fulfilled.
In this case the conformal transformations
\begin{equation}
\label{conform}
\xi_\pm \longrightarrow \widetilde\xi_\pm = A_\pm(\xi_\pm)\,,\qquad A^{\prime}\not= 0\,,
\end{equation}
allow to reduce the equation (\ref{1_sys}) to  Liouville equation; the function $\varphi(\xi^0,\xi^1)$ will be
the only dynamical variable here.
In our approach there are no  restrictions for the real functions $\rho_\mp$ which will be the dynamical variables too.
For example, the identity $\rho \equiv 0$ should  take place on any interval $[a,b]$.
We must  emphasize that in this case there are no  transformations (\ref{conform})  that reduce
the equation (\ref{1_sys}) to Liouville equation globally.

Let us point out the principal differences between Liouville equation and system (\ref{TL_rho}).
 These differences are:
\begin{enumerate}
\item The system (\ref{TL_rho}) has a ''good'' translate-invariant solution  $\varphi \equiv const $,
$\rho_\pm \equiv 0$  (classical vacuum). It should be reminded that
the corresponding solution for Liouville equation is unsatisfying
because $\varphi \equiv - \infty $.  This situation for Liouville
theory leads to some problems, in a quantum theory for example.
\item The group ${\mathcal G}$ of the  system  invariancy
is essentially  wider then  the group (\ref{conform}).
Indeed, let the functions $\varphi(\xi_+,\xi_-)$,  $\rho_\pm(\xi_\pm)$
and $\alpha_\pm(\xi_+,\xi_-)$
 be solutions of the system (\ref{TL_rho}). Then the transformation
\begin{equation}
     (\varphi, \rho_\pm,             \alpha_\pm)\longrightarrow
(\tilde\varphi, \tilde\rho_\pm, \tilde\alpha_\pm),
     \label{group_G}
     \end{equation}
     gives the new solution of the system (\ref{TL_rho}) if
     \begin{eqnarray}
  \tilde\varphi(\xi_+,\xi_-)&=&\varphi(A_+(\xi_+),A_-(\xi_-))+
     f_+(\xi_+)+f_-(\xi_-),\nonumber\\
  \tilde\rho_\pm(\xi_\pm)&=&
\rho(A_\pm(\xi_\pm))A_\pm^{\prime}(\xi_\pm)\exp{(-f_\pm(\xi_\pm))},
     \nonumber\\
\tilde\alpha_\pm(\xi_+,\xi_-)&=&
\alpha_\pm(A_+(\xi_+),A_-(\xi_-))\exp{(f_\pm(\xi_\pm))}
+g_\pm(\xi_\pm).\nonumber
     \end{eqnarray}
     for arbitrary real functions $f_\pm(\xi)$,  $g_\pm(\xi)$ and such real functions
    $A_\pm(\xi)$ where the conditions $A_-^{\prime}A_+^{\prime}\not= 0$ take place.
\item The system (\ref{TL_rho}) has a different hamiltonian structure \cite{TrLv}.
\end{enumerate}

From the geometrical point of view,
 two kinds of the transformations  (\ref{group_G}) exist. First kind ones correspond to the conformal
reparametrizations of the same world-sheet. The equalities
\begin{equation}
\label{same}
f_\pm(\xi) = -\ln A_\pm^\prime (\xi)
\end{equation}
extract these transformations from the group ${\mathcal G}$. Second kind are the rest transformations
which connect the different world-sheets.

In accordance with  point one,  we will use the following
asymptotical conditions in our theory:
\begin{eqnarray}
\label{fi_as}
\lim_{\xi^1\to\pm\infty}\varphi(\xi^1)& = & C_\pm\,,\\
\label{ro_as}
\lim_{\xi\to\pm\infty}\rho(\xi) & = & 0\,.
\end{eqnarray}

The geometrical meaning of these conditions is as follows:  we
consider the world-sheets that are planes on spatial infinities. In
spite of simplicity and naturalness of such asymptotical behavior
for a world-sheet, these conditions are unusual in a string theory.
So, the integral curvature of a world-sheet in our approach is
finite:
\begin{equation}
\label{int_cur}
{\mathcal K} = \int k dS < \infty\,.
\end{equation}
The quantity $k = \det {\bf II}/\det {\bf I}$ is a Gauss curvature
here. Note that the integral curvature will be ${\mathcal G}$ - an
invariant value. For subsequent purposes  we must reinforce the
 supposition (\ref{ro_as}). We suppose that
 the functions $\rho_\pm(\xi)$ are the functions from Shwarz space.

Let us pay attention to arbitrary constant $\varkappa$  which seemed to be an
 unnecessary value in the definition (\ref{e_pm}) before.
Taking into account the formula for the first world-sheet form
$({\bf I})$, we  justify the  introduction of the constant
$\varkappa$ by demanding $C_- =0$ in our subsequent considerations.
Thus the constant $\varkappa$ parameterize the ''classical vacua''
set for system (\ref{TL_rho}). Moreover we demand that
\begin{equation}
\label{K_lim}
   \lim_{\xi^1\to - \infty} K(\xi^0,\xi^1) = 1_2\,.
\end{equation}
That's why we restrict the transformations  (\ref{group_G}) by the condition
\begin{equation}
\label{f_lim}
\lim_{\xi^1\to - \infty}\left(f_+(\xi^1+\xi^0)+f_-(\xi^1-\xi^0)\right) = 0\,
\end{equation}
and
\begin{equation}
\label{g_lim}
\lim_{\xi^1\to - \infty} g_\pm(\xi^1\pm\xi^0) = 0\,
\end{equation}
for the functions $f_\pm$  and  $g_\pm$.

\section{Factorization of the  world-sheet set.}
\nopagebreak
\indent

Let the vectors ${\displaystyle\sf\bf b}_\mu$ be constant vectors so
that ${\displaystyle\sf\bf b}_\mu {\displaystyle\sf\bf b}_\nu =
g_{\mu\nu}$.  Let the vector - matrix ${\bf\hat E_{0}}=
{\displaystyle\sf\bf b}_0   {\boldsymbol{1_2}} - \sum_{i=1}^3
{\displaystyle\sf\bf b}_i  {\boldsymbol{\sigma_i}}$ correspond to
the basis ${\displaystyle\sf\bf b}_\mu$. It is clear that
\begin{equation}
\label{ETT}
{\bf\hat E}_\pm (\xi_\pm)  = T_\pm(\xi_\pm) {\bf\hat E_{0}} T^{\top}_\pm(\xi_\pm)\,,
\end{equation}
where $T_\pm(\xi) \in SL(2,R)$.
In accordance with formula (\ref{field_K}) the equality
\begin{equation}
\label{KTT}
K(\xi^0,\xi^1)\equiv T_+(\xi_+)T^{-1}_-(\xi_-)\,.
\end{equation}
takes place.
The following proposition can  be deduced directly from the definitions of the matrices $T_\pm$ and $K$:

{\bf Proposition 1.}
\begin{em}
The matrices $T_\pm$ are the solutions for the linear problems
\begin{equation}
     T^{\,\prime}_\pm(\xi)+
Q_\pm(\xi)T_\pm(\xi) = 0 \,,
     \label{spect1}
     \end{equation}
where
\begin{equation}
\label{Q_def}
Q_-(\xi^0,\xi^1) = K^{-1}\partial_-K\,,\quad Q_+ (\xi^0,\xi^1) =- (\partial_+K)K^{-1}\,.
\end{equation}
\end{em}

The global Lorenz transformations in our theory are the transformations
\begin{equation}
\label{E_transf}
{\bf\hat E_{0}} \longrightarrow {\widetilde{\bf\hat E}_{0}}
= B{\bf\hat E_{0}}B^{\top}\,,
\end{equation}
where the constant matrix $B \in SL(2,R)$. It is clear that these transformations correspond to
the arbitrariness for the matrix - solution of the systems (\ref{spect1}):
$$ T_\pm \longrightarrow {\widetilde T}_\pm = T_\pm B_\pm\,,\qquad B_\pm \in SL(2,R)\,,$$
where $B_+=B_-=B$ in accordance with the formulae (\ref{K_lim}) and
(\ref{KTT}).
 Thus the coefficients of the problems (\ref{spect1}) are local functions of the introduced
  variables $\varphi$, $\rho_\pm$  and $\alpha_\pm$.   These coefficients are relativistic invariants.
  For example,   the equalities
\begin{equation}
\label{Q_rho}
 Q_{12+} = -\rho_+\,,\qquad     Q_{21-} = -\rho_-\,,
\end{equation}
 will be important for our subsequent considerations.
Our  investigations  are devoted to the string excitations that
should be appropriate candidates for some anyon-type quasiparticles.
That's why we restrict the transformations (\ref{E_transf}) by
demanding $B \in SO(2)$  in the  considered model. So, we can write
that
\begin{equation}
\label{T_TU}
T_\pm(\xi) = T_{0\pm}(\xi)U(\beta)\,,
\end{equation}
where
$$ T_{0\pm}(\xi) \in SL(2,R)\,,\qquad \lim_{\xi\to - \infty}T_{0\pm}(\xi) = 1_2\,, \qquad
U(\beta)\in SO(2)\,.$$
Let  ${\mathcal G}_0$ be subgroup of the group ${\mathcal G}$ so that $A_\pm(\xi)\equiv \xi$ for all transformations
(\ref{group_G}). Then the following proposition is true:

{\bf Proposition 2.}
\begin{em}
If the group ${\mathcal G}_0$ acts on a solution \{$\varphi$,
$\rho_\pm$,  $\alpha_\pm$\} of the system (\ref{TL_rho}), the
matrices  $T_\pm$ are transformed as follows:
\begin{equation}
T_\pm\longrightarrow\tilde T_\pm={G}_\pm^{-1}T_\pm\,,
\label{T_trasf}
\end{equation}
where
$${G}_+=
\begin{pmatrix}
{\rm e}^{{f_+}/{2}}&~&0\\
~&~&~\\
g_+{\rm e}^{-{f_+}/{2}}&~&{\rm e}^{-{f_+}/{2}}
\end{pmatrix}\,,
\qquad
{G}_- =
\begin{pmatrix}
{\rm e}^{-{f_-}/{2}}&~& g_-{\rm e}^{-{f_-}/{2}}\\
~&~&~\\
0&~&{\rm e}^{{f_-}/{2}}
\end{pmatrix}\, .$$
~~\\
\end{em}

Proof is direct consequence of the formulae (\ref{gauss}),
(\ref{KTT}) and  explicit form for the transformations
(\ref{group_G}). Let ${\mathcal G}_0[T]$ denote the orbit of the
group ${\mathcal G}_0$ for matrix $T=T_\pm(\xi)$. Then the following
corollaries are fulfilled.

{\bf Corollary 1.}
\begin{em}
The unique matrix ${\mathcal U} \in {\mathcal G}_0[T]$ exists so that ${\mathcal U} \in SO(2)$.
\end{em}

Indeed, let us consider the Iwasawa decomposition for matrices
$T_\pm(\xi)$ so that $T_\pm={\mathcal E_\pm N_\pm U_\pm}$ where the
matrices ${\mathcal E}_\pm$ are diagonal matrices with positive
elements, ${\mathcal N}_+$ (${\mathcal N}_-$) -- lower (upper)
triangular matrix and ${\mathcal U}_\pm\in SO(2)$. Because of the
unique existence of such decomposition for any matrix $T \in
SL(2,R)$ and rule (\ref{T_trasf}) for the matrix $T_\pm$
transformation the proof of the Corollary 1 is clear.

{\bf Corollary 2.}
\begin{em}
The matrices ${\mathcal U}_\pm(\xi) $ satisfy the linear problems
\begin{equation}
{\mathcal U}^\prime_\pm(\xi)   +  {Q}_\pm(\xi)   {\mathcal U}_\pm(\xi) = 0\,,
\label{spect2}
\end{equation}
where matrices $Q_\pm$ are as follows: $Q_\pm =
-{\rho_\pm}\boldsymbol{\sigma_\pm} + {\rho}_\pm
\boldsymbol{\sigma_\mp}$.
\end{em}

We use same characters for  matrices $Q$ (similar for coefficients
$\rho$) both in the spectral problem (\ref{spect1})
 and in the spectral problem (\ref{spect2}); we hope that these notations don't lead to any ambiguities.

The following step is the reconstruction of the world-sheet through
the matrix elements $t_{ij\pm}$ of the matrices $T_\pm$. Taking into
account  the formula (\ref{ETT}) and the definition of the matrices
${\bf\hat E}_\pm (\xi_\pm)$, we obtain the following equalities:
\begin{equation}
  \pm   \partial_{\pm}X(\xi_\pm)
 = \frac{\varkappa}{2}\Bigl[ \left( t_{i1{\pm}}^2 + t_{i2{\pm}}^2 \right)
  \,{\displaystyle\sf\bf b}_0 -
    2\left( \,t_{i1{\pm}}{t}_{i2{\pm}}\right)\, {\displaystyle\sf\bf b}_1
   - \left(t_{i1{\pm}}^2 - t_{i2{\pm}}^2\right)\,{\displaystyle\sf\bf b}_3 \Bigr] \,,
\label{dX}
 \end{equation}

  where index  $i$  corresponds to the sign $\pm$ according to the rule     $i=\frac{3\mp1}{2}$.

  The group ${\mathcal G}_0$ can be decomposed into two kinds of the  special  transformations:
\begin{eqnarray}
(A) \quad \alpha_\pm & \to & \alpha_\pm+g_\pm \,, \nonumber\\
   \!(B)  \quad\varphi & \to & \varphi + f_+ +f_-\,,\quad \rho_\pm \to \rho_\pm e^{-f_\pm}\,,\quad
   \alpha_\pm  \to   \alpha_\pm e^{f_\pm} \,. \nonumber
   \end{eqnarray}

 The following proposition is true.

{\bf Proposition 3.}
\begin{em}
The transformation (A) does not change the world-sheet; the
transformation (B) transforms the world-sheet to the other one so
that
\begin{equation}
\label{Weyl}
 {\bf I} \longrightarrow\widetilde{\bf I} =e^{-f} {\bf I}\,,\qquad
 f = f_+(\xi_+) +f_-(\xi_-)\,,
\end{equation}
and the second form {\bf II}:
\begin{equation}
\label{form_II}
{\bf II}   \longrightarrow\widetilde{\bf II}
= \varkappa [\rho_+ e^{-f_+}d\xi^{2}_+   -  \rho_- e^{-f_-}d\xi^{2}_- ]
\end{equation}
\end{em}
 The proof follows from the explicit formulae (\ref{dX}) for tangent vectors $\partial_{\pm}X(\xi_\pm)$,
explicit formulae for the forms {\bf I} and {\bf II} and rules
(\ref{T_trasf}) for transformations of the matrix elements
$t_{ij\pm}$. As a comment we note that the existence of the
transformations (A), which do not change the world-sheet, is the
consequence of the arbitrariness in the definition of matrices
${\bf\hat E}_\pm (\xi_\pm)$.

Let ${\mathcal X}$ be the set of a world-sheets introduced in the beginning of the paper.
The object of our subsequent investigations is factor-set ${\mathcal X}/{\mathcal G}$ only.
Taking into account   Corollary 1 we can choose the representatives in every coset so that
$T_\pm = {\mathcal U}_\pm \in SO(2)$. In accordance with the formula (\ref{T_TU})
we have
\begin{equation}
\label{UU}
{\mathcal U}_\pm(\xi) = {\mathcal U}_{0\pm}(\xi)U(\beta)\,,
\qquad
\lim_{\xi\to - \infty}{\mathcal U}_{0\pm}(\xi) = 1_2\,,
\end{equation}
where the matrix $ {\mathcal U}_{0\pm}(\xi) \in SO(2)$.   Because of
the boundary conditions on the matrices ${\mathcal U}_{0\pm}$ the
one-to-one correspondence $\rho_\pm(\xi) \leftrightarrow {\mathcal
U}_{0\pm}(\xi)$ exists for each sign $\pm$.
 Let us introduce the quantities
 $I_\pm(\xi) = \int_{-\infty}^\xi\rho_\pm(\eta)d\eta\,. $
 It is easy to see that
 \begin{equation}
\label{matr_U}
 {\mathcal U}_{0\pm}(\xi) =
 \left(\begin{matrix}
   \cos I_\pm(\xi)& \pm \sin I_\pm(\xi)\\
\mp \sin I_\pm(\xi)& \cos I_\pm(\xi)
\end{matrix}\right)\,.
  \end{equation}
 In the context of the factorization procedure, as defined above,
  we are able to write the principal minor $K_{11}$ of the matrix $K(\xi^0,\xi^1)$
  as the function of the quantities $\rho_\pm$. To do it we must extract the element $K_{11}$
 from the formula (\ref{gauss}). The result is as follows:
 \begin{equation}
\label{phi_rho}
 \exp[-\varphi(\xi^0,\xi^1)] = \cos^2[I_+(\xi_+) +I_-(\xi_-)]\,.
 \end{equation}
 This equality can be considered as the  geometrical gauge condition for our theory.
 It must be emphasized that the arbitrariness (\ref{conform}) is not fixed anywhere.

 The formulae for tangent vectors $\partial_{\pm}{\bf X}(\xi_\pm)$ for chosen representatives
 is as follows:
  \begin{equation}
\label{tangent}
  \pm \partial_{\pm}{\bf X}(\xi_\pm)  =  \frac{\varkappa}{2}\Bigl[ {\displaystyle\sf\bf b}_0
  \mp {\displaystyle\sf\bf n}(\beta)\Bigr]  + \varkappa \delta{\bf e}_\pm(\xi_\pm)\,,
  \end{equation}
  where
  $$ {\displaystyle\sf\bf n}(\beta) =(\sin 2\beta) {\displaystyle\sf\bf b}_1 + (\cos 2\beta) {\displaystyle\sf\bf b}_3 $$
  and
  $$    \delta{\bf e}_\pm(\xi;\beta)=  \Bigl(\sin I_\pm(\xi)\Bigr)
  \Bigl[ - \cos(I_\pm(\xi) \pm 2\beta) {\displaystyle\sf\bf b}_1
  \pm \sin (I_\pm(\xi) \pm 2\beta){\displaystyle\sf\bf b}_3\Bigr]  \,.$$
 It is clear that
  $$ X_0(\xi^0,\xi^1) =  \varkappa\xi^0 {\displaystyle\sf\bf b}_0 + Z_0\, \qquad   (Z_0 = const) \,,$$
  for our gauge.

 To reconstruct the spatial coordinates ${X}_j(\xi^0,\xi^1)$  ($j=1,3$) of the world-sheet  through the derivatives,  we must add two-dimensional vector with   components $Z_1$ and $Z_3$.
 Taking into account  our subsequent definition of the constants  $Z_j$, we suppose that
    the following conditions were fulfilled:
  \begin{equation}
  \label{topol}
  I_\pm(+\infty) = \int_{-\infty}^\infty \rho_\pm(\eta)d\eta = \pi n_\pm\,,
  \end{equation}
  where $n_\pm$ are integer numbers. These conditions will be very important for our theory.
  Later we shall state the topological nature of the numbers $n_\pm$.

  Let  the functions $s_{j\pm}(\omega;\beta)$ be Fourier - transformations for
  the components of the vectors $\delta{\bf e}_\pm(\xi;\beta)$:

  $$ s_{j\pm}(\omega;\beta) =
  \frac{1}{2\pi}\int_{-\infty}^\infty e^{-i\xi\omega}(\delta{\bf e}_\pm(\xi;\beta))_j\,   d\xi \,,
  \qquad  j= 1,3\,.$$
  The Shwarz functions $\rho_\pm$ are subjected to the conditions (\ref{topol}), that is why
  the components $(\delta{\bf e}_\pm(\xi;\cdot))_j$  will be  Shwarz functions. Thus
   the components $s_{j\pm}(\omega;\cdot)$ will be well-defined Shwarz functions too.
   Taking into account the restriction of the space-time invariancy group
 from $E(1,2)$ to $E(2)$, the final formula for spatial string coordinates $X_j$
 will be as follows ($j=1,3$):
 \begin{eqnarray}
 \label{X_final}
   X_j(\xi^0,\xi^1) &=& Z_j -\varkappa{\displaystyle\sf\bf n}(\beta)\xi^1 + \\[3mm]
  ~&+& i\varkappa \,V.p.\int_{-\infty}^\infty \frac{s_{j+}(\omega;\beta)}{\omega}\,e^{i\xi_+\omega} d\omega
  - i\varkappa \,V.p.\int_{-\infty}^\infty \frac{s_{j-}(\omega;\beta)}{\omega}\,e^{i\xi_-\omega} d\omega\,.\nonumber
 \end{eqnarray}

 Thus we have the following one-to-one correspondence:
  \begin{equation}
\label{corresp}
 \Bigl(   X_1(\xi^0,\xi^1),\, X_3(\xi^0,\xi^1) \Bigr) \longleftrightarrow
 \Bigl( \rho_+(\xi_+), \rho_-(\xi_-)\,;\, Z_1, Z_3,  \beta,  \varkappa \Bigr)\,.
 \end{equation}
 The ''external'' variables $( Z_1, Z_3,  \beta,  \varkappa )$ correspond to the
 translations, rotations and scale transformations of the considered strings.
 The following section is devoted to hamiltonian structure of suggested theory.

 \section{Hamiltonian structure.}
\indent

Let us write the formulae for  N\"oether invariants of the action
(\ref{action}):
$$P_\mu = \gamma \int_{-\infty}^\infty \partial_0 X_\mu\, d\xi^1\,,\qquad
M_{\mu\nu} = \gamma \int_{-\infty}^\infty \left(\partial_0 X_\mu
X_\nu  -  \partial_0 X_\nu  X_\mu\right) d\xi^1\,. $$ These
expressions are formal unless the  convergence of the integrals will
be stated. The only quantities that  are interesting  in our case
are spatial momentum ${\bf P} = {\displaystyle\sf\bf b}_1 P_1 +
{\displaystyle\sf\bf b}_3 P_3 $ and the component $M = M_{13}$\,;
that is why we investigate the convergence of these quantities only.

{\bf Proposition 4.}
\begin{em}
The quantities $P_1$, $P_3$  and $M$ are finite if the conditions
(\ref{topol}) are fulfilled.
\end{em}

Indeed, formulae  (\ref{tangent}) and   (\ref{X_final}) lead to the following expressions:
\begin{eqnarray}
\label{P_1}
P_1 &=&  - \gamma\varkappa\int_{-\infty}^\infty\left[ \sin(2I_+(\xi)+2\beta) + \sin(2I_-(\xi)-2\beta) \right]\,d\xi\,, \\[3mm]
P_3 &=& - \gamma\varkappa\int_{-\infty}^\infty\left[ \cos(2I_+(\xi)+2\beta) - \cos(2I_-(\xi)-2\beta)  \right]\,d\xi\,,
\label{P_3}
\end{eqnarray}
and
$ M = Z_1 P_3 - Z_3 P_1 +J$, where
\begin{eqnarray}
\label{J}
J &=& \gamma\varkappa^2 F_J(\rho_+,\rho_-) \equiv
\gamma\varkappa^2 \int\limits_{-\infty}^\infty \int\limits_{-\infty}^\infty \varepsilon(\eta_1-\eta_2)
\sin I_+(\eta_1)\sin I_-(\eta_2)\, \times \nonumber\\[3mm]
~& \times & [\cos I_+(\eta_1)\sin I_+(\eta_2) + \cos I_-(\eta_1)\sin I_-(\eta_2)]\,d\eta_1 d\eta_2\,.
\end{eqnarray}
The '' constraints'' (\ref{topol}) were the sufficient conditions for the convergence of integrals in right hand sides
of the formulae (\ref{P_1}), (\ref{P_3}) and (\ref{J}). This Proposition confirms the topological nature of the
conditions (\ref{topol}): there is no any physical theory outside of ''constraint surface'' (\ref{topol}).

{\bf Proposition 5.}
\begin{em}
The quantities $P_1$, $P_3$, $J$, $\rho_\pm$  are constrained by the
following condition:
\begin{equation}
\label{constr_1}
 \Phi(\rho_+,\rho_-, \,{\bf P}, J ) \equiv
    {\bf P}^2  - \gamma J  \Omega(\rho_+,\rho_-) = 0\,,
\end{equation}
where  $\Omega(\rho_+,\rho_-) = F_J^{-1}(\rho_+,\rho_-) F_P(\rho_+,\rho_-) $ and
\begin{eqnarray}
F_P(\rho_+,\rho_-) &=&
\left[\int_{-\infty}^\infty [ \sin(2I_+(\xi)+2\beta) + \sin(2I_-(\xi)-2\beta) ]\,d\xi\right]^2 +\,, \nonumber\\[3mm]
~ & + & \left[\int_{-\infty}^\infty[ \cos(2I_+(\xi)+2\beta) - \cos(2I_-(\xi)-2\beta)  ]\,d\xi\right]^2 = \nonumber\\[3mm]
~ &=& 4\int_{-\infty}^\infty \int_{-\infty}^\infty \sin(I_+(\xi) + I_-(\xi))\sin(I_+(\eta) + I_-(\eta))\times
\nonumber\\[3mm]
~& \times & \cos\Bigl(I_+(\xi) - I_-(\xi) - I_+(\eta) + I_-(\eta)\Bigr)\, d\xi d\eta \,. \nonumber
\end{eqnarray}
\end{em}

The proof is the exclusion of the constant $\varkappa$ from the formulae  (\ref{P_1}), (\ref{P_3}) and (\ref{J}).

We choose the constants $(P_1, P_3, J)$  as the external variables in our theory instead the constants $(\beta, \varkappa)$.
Since we introduce three variables instead two ones,
the condition (\ref{constr_1}) must be imposed as  the constraint.
In accordance with our supposition the constant $\varkappa$  is non-zero finite constant; for this domain
the identities $\rho_\pm\equiv 0$ lead to the equalities $|{\bf P}| =0$, $J=0$.
  We extend  considered dynamical system now: the inequalities $|{\bf P}| \not=0$ and (or)
$J \not=0$ will be admitted in spite of  the identities
$\rho_\pm\equiv 0$ hold.
Moreover the equalities $|{\bf P}| =0$, $J=0$ will be admitted for
arbitrary functions $\rho_\pm$.
 This extension corresponds to the addition of the boundary points
$\varkappa = 0$ and $\varkappa = \infty$ to the primary domain $(0,\infty)$.

Thus the hamiltonian dynamical system in our theory is defined as follows:
\begin{itemize}
\item
the hamiltonian variables
$$ \Bigl( \rho_+(\xi_+), \rho_-(\xi_-)\,;\, Z_1, Z_3, P_1, P_3,\, J\Bigr)\,; $$
\item
Poisson brackets
\begin{eqnarray}
\label{br_rho}
\{\rho_\pm(\xi),  \rho_\pm(\eta)\} & = & \mp\delta^\prime(\xi - \eta)\,, \\[3mm]
\{P_i,Z_j\} &=& \delta_{ij} \,
\end{eqnarray}
(other possible brackets are equal to zero);
\item
constraints (\ref{topol}) and (\ref{constr_1});
\item
hamiltonian
$$ H = \frac{1}{2}\int_{-\infty}^\infty \left[\rho_+^2(\xi) + \rho_-^2(\xi)\right]\,d\xi +
l(\xi^0)\Phi\,,$$ where the function $l$ is a lagrange multiplier.
\end{itemize}

Note that dynamical variable $J$ annihilates all Poisson
brackets\,\footnote{This definition is motivated by the fact that
the value $J$ in non-relativistic case is the only non-zero
component of Luban'sky - Pauli vector.}. The phase space will be as
follows:
$$ {\mathcal H} = {\mathcal H}_+ \times {\mathcal H}_- \times {\mathcal H}_p\,,$$
where  ${\mathcal H}_\pm$ are the phase spaces of free
two-dimensional fields $\rho_\pm$ and ${\mathcal H}_p$ is the phase
space of a free particle on a plane. The model is non-trivial
because of the constraint (\ref{constr_1}), which entangles the
internal and external  variables. This constraint selects the
''Regge trajectories'' $J = \alpha^{\prime} {\bf P}^2$, where the
slope $\alpha^\prime$ depends on the internal variables only.
 Topological constraints (\ref{topol}) select the symplectic sheets in the spaces ${\mathcal H}_\pm$.

 In the next section we investigate the degrees of freedom which correspond to
 ''internal'' variables $ (\rho_+(\xi_+), \rho_-(\xi_-))$.

 \section{Spectral problems and N - soliton strings}
 \indent

 Let us consider the pair of spectral problems

 \begin{equation}
    {\mathcal U}^{\,\prime}_\pm(\xi^0,\xi^1;\lambda)+
Q_\pm(\xi_\pm){\mathcal U}_{\pm}(\xi^0,\xi^1;\lambda) =
{\rm i}\lambda {\boldsymbol{\sigma_3}}{\mathcal U}_{\pm} (\xi^0,\xi^1;\lambda) \,.
     \label{sp_task}
     \end{equation}

 These spectral problems are similar to the ones for Non-linear Shroedinger equation,  so we shall use the results of \cite{TakFad}.
  Because the functions $\rho_\pm = \rho_\pm(\xi^1\pm \xi^0) $ are the functions from Shwarz space,
     we can define the matrices ${\mathcal U}^{[\pm]}(\xi^0,\xi^1;\lambda)$ by means of the asymptotical conditions
  \begin{equation}
     \label{as_1}
{\mathcal U}^{[\pm]}(\xi^0,\xi^1;\lambda)
 \smash{\mathop{\longrightarrow}\limits_{\xi^1\to\pm\infty}}
 \exp({\rm  i}\lambda\xi^1{\boldsymbol{\sigma_3}})\,.
    \end{equation}
 We omit the low indices $\pm$, which mark the spectral problems (\ref{sp_task}), in the formula (\ref{as_1}) as
 well as in all the following formulae, where such notations don't lead to any ambiguities.
 The monodromy martices $\mathfrak{M}(\lambda)$
 $$\mathfrak{M}_\pm(\lambda)=
\left(\begin{matrix} {\mathfrak a}_\pm(\lambda)& {\mathfrak b}_\pm(\lambda)\\
     -\overline{\mathfrak b}_\pm(\lambda)& \overline{\mathfrak a}_\pm(\lambda)
\end{matrix}\right)\,$$
  are defined as follows:
 \begin{equation}
     {\mathcal U}^{[+]}(\xi;\lambda)={\mathcal U}^{[-]}(\xi;\lambda )\mathfrak{M}( \lambda)\,.
     \label{monodr}
     \end{equation}
   The  conditions (\ref{topol}) mean that
   \begin{equation}
   \label{M_1}
   \mathfrak{M}_\pm(0) = (-1)^{n_\pm}\,.
   \end{equation}
 Because the group $SO(2)$ is  infinitely connected group, the numbers  $n_\pm$ have topological nature.

 There exists a one - to - one correspondence between the coefficients $\rho_\pm$ and the scattering
 data  of the spectral  problems  (\ref{sp_task}) \cite{TakFad}.
 Below in this section  we consider the N - soliton case only so that the equalities
 $$ {\mathfrak b}_\pm(\lambda) \equiv 0\,$$
 take place. This supposition means that the functions $\rho$  can be reconstructed through
 the discrete spectrum $\lambda_n$, $c_n$, where $n=n_\pm =1,\dots, N_\pm$. The constants $\lambda_n$ are the
 eigenvalues of the spectral problems (\ref{sp_task}); the  additional constants $c_n$  are the proportionality coefficients
 for the appropriate columns of the matrices ${\mathcal U}^{[+]}(\lambda)$ and ${\mathcal U}^{[-]}(\lambda)$ at $\lambda = \lambda_n$.
 The reality condition for the functions $\rho$  means that the spectrum $\{\lambda_n\}$
 consists of
 pure imaginary points as (or) a symmetrical pairs $( \lambda_n, \lambda_m)$ so that  $\lambda_n = - {\overline\lambda}_m$; there are no any other kinds of the values ${\lambda_n}$
 in the spectrum.
  The reduction to N - soliton case means that the functions ${\mathfrak a}_\pm(\lambda)$ are integer in a complex
 $\lambda$ - plane so that the matrices  ${\mathcal U}^{[\pm]}(\xi;\lambda)$ can be reconstructed explicitly
 by means of the matrix Riemann problem \cite{TakFad}. Because
    ${\mathcal U}_{0\pm}(\xi) ={\mathcal U}^{[-]}_{\pm}(0,\xi;0)$,  the world-sheet
  can be reconstructed through  the scattering data of the systems (\ref{sp_task}) too.
  Note that conditions (\ref{M_1}) are fulfilled here.
   For N -soliton case, the numbers $n_\pm$ will be the numbers
 of pure imaginary points in the spectrum of the corresponding spectral problem.

\begin{figure}[h]
\centering\includegraphics[width=2.0in,angle=-90]{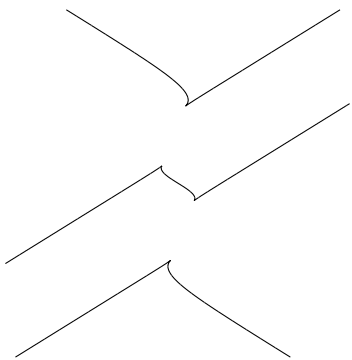}
\parbox[t]{10cm}{\caption{World-lines of the cuspidal points: example for 3-soliton case.}
\label{3sol_braid}}
\end{figure}

  Let us consider some examples.
 \begin{itemize}
 \item  $N_+ = N_- =0$. It is clear that  ${\mathcal U}_{0\pm}(\xi) = 1_2$ in this case so that
 the string is a straight line and the world-sheet is a time-like plane.
 \item  $N_+ = N_- =1$,  $\lambda_{1\pm} = {\rm i}a_\pm$, $a_\pm$ -- positive real numbers.
 The reconstruction of the matrix elements $ u_{ij} = ({\mathcal U}_0)_{ij}$ leads to the
 explicit formulae
$$ u_{11}(\xi)= \frac{e^{-4 a \xi}-c^2}{e^{-4 a \xi}+c^2} \,\,,\qquad
u_{12}(\xi)= -\frac{2c e^{-2a\xi}}{e^{-4 a\xi}+c^2}\,\,.$$ The
details can be founded in the work \cite{TrLv} where similar
calculations have been carried out in accordance with the method
\cite{ZMNP}. The string has two moving cusps in this case (see
\cite{TMF07}, where the pictures of both the string and the world -
sheet are presented for this case).

 The world-lines of the cusps form the braid in the space-time $E_{1,2}$ which is
 demonstrated\,\footnote{The axis $X_0$ in $E_{1,2}$ corresponds to vertical direction on the Figures \ref{2sol_braid},
 \ref{4sol_braid} and  \ref{3sol_braid}. }
 in Fig.\ref{2sol_braid}.    Of course, such picture of the cusp world-lines  corresponds
 to the self-intersected world-sheet.

\begin{figure}[ht]
\centering\includegraphics[width=2.0in,angle=-90]{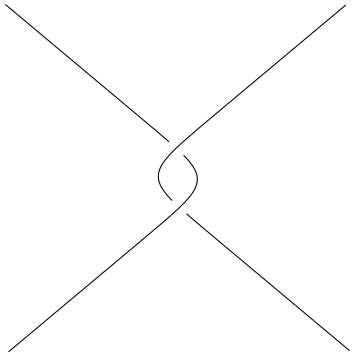}
\parbox[t]{10cm}{\caption{The braid in 2-soliton case.}
\label{2sol_braid}}
\end{figure}

 \item  $N_+ = N_- =2$,  $\lambda_{j\pm} = {\rm i}a_{j\pm}$, $a_{j\pm}$ -- arbitrary positive real numbers.
 The example of corresponding braid is demonstrated
 in Fig. \ref{4sol_braid}. It should be stressed that braid topology depends on the values of the constants
 $\lambda_n$, $c_n$. So, the intersection of some world-lines is possible for some
 conditions
 \begin{equation}
 \label{inters}
 F(\lambda_1,\dots,\lambda_N; c_1,\dots,c_N) =0\,,
 \end{equation}
 connected with the discrete spectrum.
 \begin{figure}[ht]
\centering\includegraphics[width=2.0in,angle=-90]{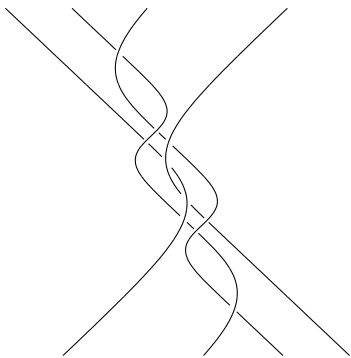}
\parbox[t]{10cm}{\caption{The example of braid in 4-soliton case.}
\label{4sol_braid}}
\end{figure}
 \end{itemize}

 The inequality $\rho_+\rho_- > 0$ was fulfilled for all the  examples considered above.
 The following Proposition is true.

  {\bf Proposition 6.}
\begin{em}
If the inequality $\rho_+\rho_- > 0$ takes place, the number of
cuspidal points on a string is a dynamical invariant.
 \end{em}

\begin{figure}[ht]
\centering\includegraphics[width=4.3in]{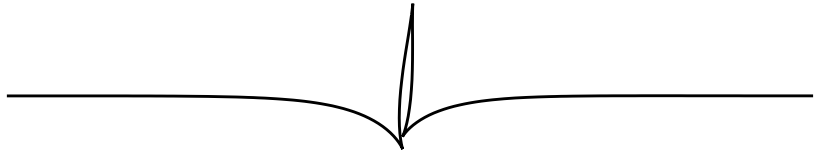}
\parbox[t]{10cm}{\caption{The string with three cuspidal points: a moment before coincidence.}
\label{3s_int}}
\end{figure}

 Indeed, the explicit formula for the  world-sheet  form {\bf I} means that the cuspidal points
 correspond to the singularities of the function $\varphi(\xi^0,\xi^1)$.
 If the inequality $\rho_+\rho_- > 0$ takes place, the equation (\ref{1_sys}) is conformally equivalent
 to the Liouvile equation. The dynamics of singularity of the solutions of Liouville equation has been
 investigated in the work \cite{JPP}. It was stated that the number of singular points is conserved by
 $\xi^0$ - dynamics and the lines of the singularities don't intersect.
  In spite of this fact  the world-lines of the cuspidal points can intersect  because the string
 with cuspidal points  can have the intersections (see Fig.\ref{3s_int} as some example)

As it is stressed above, the suggested approach admits any sign of
the product  $\rho_+\rho_- $.
 The picture of the  world-lines is more complicated when the  condition $\rho_+\rho_- > 0$ was
 disturbed. Corresponding example for $N_+ = 2$,  $N_- =1$ is demonstrated on Fig.\ref{3sol_braid}

  Thus   we  have the pictorial model for classical quasiparticles
  which can interact, be created or  annihilated on a plane.

{\bf Acknowledgments.} Author thank V. S. Looze for computer visualization of considered examples.


\begin{thebibliography}{99}
\bibitem{Kitaev}{A. Yu. Kitaev // Annals Phys. 2003, ({\bf 303})  p. 2 - 30. ArXiv:   quant-ph/9707021}
\bibitem{LevWen}{M. Levin, X.-G. Wen // Rev. Mod. Phys. V. 77 p. 871. 2005. ArXiv:  cond-mat/0407140v2.}
\bibitem{Zwi}{B. Zwiebach // A First course in String Theory. Cambr. Univ. Press (2004);}
\bibitem{GShW}{M. Green, J.  Shwarz and E. Witten. Superstring theory. v.1. Cambrige Univ. Press. 1987.}
\bibitem{TMF00}{S.V. Talalov // Theor. Math. Phys., 2000, ({\bf 123}):1, 446 -- 450;
{ Journ. of Phys. A}. V. 22.  P. 2275. 1989. }
\bibitem{BarNes}{B.M. Barbashov and V.V. Nesterenko. Introduction to the Relativistic String Theory.
World Scientific, Teaneck, N.J. (1990);}
\bibitem{TMF07}{S.V. Talalov // Theoretical and Mathematical Physics, 2007, ({\bf 152}): 3, 1234 -- 1242.}
\bibitem{TMF09}{S.V. Talalov // Theoretical and Mathematical Physics, 2009, ({\bf 159}): 1, 469 -- 473.}
\bibitem{TrLv} { S.V. Talalov  // Theoretical and Mathematical Physics, 1987, ({\bf 71}):3, 588 -- 597.}
\bibitem{TakFad}{L.A. Takhtajan, L.D. Faddeev. Hamiltonian Approach to Soliton Theory. Nauka: Moscow, 1986
(in Russian)}
\bibitem{ZMNP}{V.E. Zakharov, S.V,  Manakov, S.P. Novikov and  L.P. Pitayevsky. The Soliton Theory: Inverse Scattering
Transform Method. Moscow: Nauka, 1980 (in Russian)}
\bibitem{JPP}{G.P. Dzhordzhadze, A.K. Pogrebkov and M.C. Polivanov // Theor. Math. Phys., 1979, ({\bf 40}), 706 -- 715}
\end{thebibliography}
\end{document}